# THE INVERSE PROBLEM OF REMOTE NEUTRINO DIAGNOSTICS OF INTRAREACTOR PROCESSES

*V.D. Rusov [a], T.N. Zelentsova, V.A. Tarasov, D.A. Litvinov*

*Department of Theoretical and Experimental Nuclear Physics,*

*Odessa National Polytechnic University, 1, Shevchenko av., 65044, Odessa, Ukraine*

Using the well-known experimental data the inverse problem of neutrino diagnostics of reactor core is considered. The solution of this problem makes it possible to determine distantly the current value of nuclear density of each nuclear fuel components with known accuracy and also opens up the possibility for the development of neutrino on-line technology of temporal evolution of nuclear fuel isotopic structure and reactor power

---

[a] Author to whom correspondence should be addressed electronic mail: siiis@te.net.ua

# I. INTRODUCTION

It is well known that the nuclear reactor is a pure and power source of an electron antineutrinos $\widetilde{\nu}_e$, whose spectrum is formed as a result of fission fragments β-decay of four isotopes of nuclear fuel: $^{235}$U, $^{239}$Pu, $^{238}$U and $^{241}$Pu. Due to its physical properties antineutrino has unique penetrating power. This allows to avoid the distorting medium effect and to detect neutrinos, which are practically identical to neutrinos produced by actinoid fission, independently of source-to-detector distance. For their detection the inverse β-decay reaction is traditionally used:

$$\widetilde{\nu}_e + p \rightarrow e^+ + n. \qquad (1)$$

The intensity of the detected neutrino events $n_\nu$ is connected with reactor thermal power $<W_{NPP}>$ by relation:

$$n_\nu = \frac{<W_{NPP}>}{<E_f>} \times \frac{\gamma\varepsilon_0}{4\pi<R>^2} \times N_p \Sigma_\nu, \quad s^{-1}, \qquad (2)$$

where $\Sigma_\nu = M_\nu <\sigma_{\nu p}>, \quad M_\nu = \int_{E_{thresh}}^{E_{max}} \rho(E_\nu) dE_\nu, \quad <\sigma_{\nu p}> = \dfrac{\int_{E_{thresh}}^{E_{max}} \sigma_{\nu p}(E_\nu)\rho(E_\nu)dE_\nu}{\int_{E_{thresh}}^{E_{max}} \rho(E_\nu)dE_\nu}. \qquad (3)$

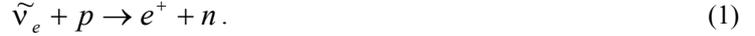

Here $<E_f>=\Sigma(\alpha_i \cdot E_{fi})$ is an average energy per fission at the given fuel composition; $\alpha_i$ is the contribution from $i$-th isotope ($i=5; 9; 8; 1$) to total fission cross-section (which depends on the method of spectrum $\rho(E_\nu)$ determination [1-9]); $(4\pi<R>^2)^{-1}$ is effective solid angle with allowance for real energy-release distribution in the core volume; $N_p$ and $\gamma\varepsilon_0$ are the detector characteristics (hydrogen atom number in the target and the detection efficiency with allowance for the part of detected neutrons γ corresponding to reaction (1)); $\Sigma_\nu$ and $<\sigma_{\nu p}>$ are neutrino reaction cross-sections (cm$^2$/fission and cm$^2$/ν-particle respectively); $\Sigma_\nu = \Sigma(\alpha_i \Sigma_{\nu i})$ at given fuel composition; $M_\nu$ is electron antineutrino number per fission; $\rho(E_\nu)=\Sigma(\alpha_i \rho_i)$ is antineutrino energy spectrum (MeV$^{-1}$× fission$^{-1}$) emitted by fission products of all fuel components; $\sigma_{\nu p}(E_\nu)$ is the interaction cross-section of monoenergetic (with an energy $E_\nu$) antineutrinos with allowance for recoil, weak magnetism and radiation corrections [10-13].

It is easy to show [14] that the basic balance equation of reactor antineutrino spectrometry, which describes the contribution $\alpha_i$ from each actinoid to experimentally measured energy spectrum $\eta(E_\nu)$, can be obtained from Eq. (2):

$$\eta(E_\nu) = \frac{\gamma\varepsilon_0 N_p}{4\pi<R>^2} \sum_i \rho_i \sigma_{\nu p}(E_\nu) \lambda_i \Delta t, \quad MeV^{-1}, \qquad (4)$$

where

$$n_\nu = \frac{1}{\Delta t}\int_{E_{thresh}}^{E} \eta(E_\nu) dE_\nu, \quad \lambda_i = \alpha_i \lambda, \quad \lambda = \frac{<W_{NPP}>}{E_f}; \qquad (5)$$

$\Delta t$ is the measurement time; $\lambda$ and $\lambda_i$ are the average and partial ($i=5; 9; 8; 1$) fission rates of nuclear fuel.

It is obvious that the obtaining of an effective solution of Eq. (4) relative to $\lambda_i$, (which along with the seeking of optimal numerical methods should also include physical substantiation of the methods and procedures for the determination of basic energy functions $\eta(E_\nu)$, $\rho_{\nu i}$, $\sigma_{\nu p}(E_\nu)$ in this equation) is the midpoint for the development of on-line neutrino method.

## II. FORMULATATION OF THE PROBLEM

Despite the abundance of the publications devoted to the detection of reactor antineutrinos, we have found only three experiments [4,15-16] executed at Rovno NPP (Ukraine) in 1984-1986, 1990 and 1997, which reflect the essence of basic equation (4) of reactor neutrino spectrometry in full measure . The positron energy spectra $S_{e+}(E_{e+})$ for reaction Eq.(1) are presented in Fig. 1-3 [4, 15, 16]. The main characteristics of experiment geometry, detector and reactor are presented in Table I.

Let us consider the obtaining of an energy spectrum $\eta(E_\nu)$ or, in other words, the spectrum "on detection place". The antineutrino spectrum can be determined directly from positron kinetic energy $T_{e+}$ as it differs from the detected neutrino energy only on threshold energy 1.804 MeV. However, due to the partial absorption of annihilation quanta in useful capacity of spectrometer observable positron energy always will be $\Delta T_{e+}$ (~ 0.6 MeV) higher than true kinetic energy, and it is necessary to take it into account when energy spectrum $\eta(E_\nu)$ is obtained [4, 15, 16]:

$$E_\nu = (T_{e+} - \Delta T_{e+}) + 1.804 MeV + O\left(\frac{E_\nu}{M_n}\right). \tag{6}$$

The numerical values of additional shift energy $\Delta T_{e+}$ are presented in Table 1 and the energy spectra $\eta(E_\nu)$ in the 2-9 MeV range obtained by Eq. (6) are given in Table II.

Now let us consider the determination of spectra $\rho_i$ "on birthplace", which differ not only "in place" but also in the way of their determination. The obtaining of calculated spectrum $\rho_i$ for given fissionable nucleus is based on the fact that ration of $\nu$-spectrum $(\rho_i)$ to $\beta$-spectrum $(\rho_\beta)$ of fission products is stationary values in secular equilibrium condition and does not depend on decay scheme [5, 17]. The constancy of this ration has no clear physical substantiation[7], nevertheless experimentally measured $\beta$-spectra of fission fragments of $^{235}$U, $^{239}$Pu and $^{241}$Pu [9, 16, 18] coincide with calculation results [5,17] with a good accuracy. The converted in this way spectra $\rho_i$, which are necessary for the further calculations, are given in Table 2. For $^{235}$U, whose $\beta$-spectrum was not measured, the calculated spectrum [5] is used.

The interaction cross-section of monoenergetic antineutrinos $\sigma_{\nu p}(E_\nu)$ is the most theoretically grounded value for the reaction (1) [10-12] among three energy functions $\eta(E_\nu)$, $\rho_i$ and $\sigma_{\nu p}(E_\nu)$. The antineutrino-capture cross-section with the allowance for its behavior close to reaction threshold ($\delta_{thr}$), the recoil ($\delta_{rec}$), weak magnetism ($\delta_{WM}$) and radiation corrections ($\delta_{rad}$) looks like:

$$\sigma_{\nu p}(E_\nu) = \sigma_0(E_\nu) \times (1 + \delta_{thr}) \times (1 + \delta_{WM} + \delta_{rec}) \times (1 + \delta_{rad}). \tag{7}$$

The analytical expressions for all corrections and their detailed discussion are given in Refs. [11, 12]. The "naive" cross-section $\sigma_0(E_\nu)$ [11, 12] corresponding to the approximation of infinitely heavy nucleons or $E_n \approx m_n$, $E_{e+} << m_n$ ($\hbar=c=1$) has form

$$\sigma_0(E_\nu) = \frac{2\pi^2 \hbar^3 \ln 2}{m_e^5 c^7 (f\tau_{1/2})} \times \frac{1}{c} \left[(E_\nu - \Delta)^2 - m_e^2 c^4\right]^{1/2} \times (E_\nu - \Delta), \tag{8}$$

where $E_\nu - (m_n - m_p)c^2 = E_\nu - \Delta = E_{e+}$ is total positron energy in the reaction (1) as it is possible to neglect the neutron recoil energy at the antineutrino energies produced in the reactor; $f\tau_{1/2}$ is so-called reduced neutron half-life [7], for which phase space factor of neutron $f=1.7146$ is determined to 0.01 % [13] and half-life is $\tau_{1/2} = \tau \ln 2$ (where $\tau=887.4 \pm 0.2$ s [19]).

If the energy and the mass are expressed as MeV in Eq. (8), we have:

$$\frac{2\pi^2 \ln 2}{m_{e^+}^5 (f\tau_{1/2})} = 9.4215 \times 10^{-44} \quad cm^2/MeV^2.$$

Now we are ready to solve Eq. (4), which move to the system of linear equations after energy discretization. The consequences of Eq. (4) discretization are visible in Table II. It is obvious that the common determinant of the system of equations has many "zero" and the system of equations as a whole can become quasi-degenerated. It means that the system of linear equations is an ill-conditioned problem and its solutions may be unstable to small changes of input data. In other words, the problem of this type belongs to ill-posed problem class and we used the Tikhonov regularizing method [20, 21] for its solution.

### III. DYNAMICS RECONSTRUCTION OF ISOTOPIC COMPOSITION OF NUCLEAR FUEL

Using well-known experimental results [4, 15, 16] let us show the efficiency of the solution of inverse problem of neutrino spectrometry (described by Eq. (4)) for intrareactor processes. For that we present input Eq. (4) as

$$y = \sum_{i=1}^{n} \lambda_i x^{(i)},$$

where

$$y = \eta(E_\nu), \quad x^{(i)} = \frac{\gamma \varepsilon_0 N_p}{4\pi R^2} \Delta t \sigma_{\nu p}(E_\nu) \rho_i$$

or after discretization:

$$y_k = \sum_{i=1}^{n} \lambda_i x_k^{(i)}, \quad k = 1, 2, ..., N,$$

where $N$ is number of energy values, at which spectrum was measured.

At this stage $\lambda_i$ is usually obtained by least-squares method from condition for the discrepancy minimum:

$$\chi = \sum_{k=1}^{N} \left( \sum_{i=1}^{n} \lambda_i x_k^{(i)} - y_k \right)^2, \tag{9}$$

i.e. at $\partial \chi / \partial \lambda_i = 0$. This reduces to the following system of linear algebraic equations relative to $\lambda_i$:

$$\sum_{i=1}^{n} \sum_{k=1}^{N} \lambda_i x_k^{(i)} x_k^{(j)} = \sum_{k=1}^{N} \sum_{i=1}^{n} x_k^{(i)} y_k, \quad j = 1, 2, ..., n,$$

which may be shortly written down as

$$A\lambda = u, \tag{10}$$

where

$$A_{ij} = \sum_{k=1}^{N} x_k^{(i)} x_k^{(j)}, \quad u_j = \sum_{k=1}^{N} x_k^{(j)} y_k$$

Let $A_{ij}$ and $u_j$ are given with the errors represented by Euclidean norm $\|A_h - A\| \leq h$ и $\|u_\delta - u\| \leq \delta$. Then according to Tikhonov regularizing method [20] the search of normal solution of Eq. (4) (or the system of equations (10)) is reduced to

the determination of vector of minimal norm on set of vectors $\lambda=(\lambda_1,\ldots\lambda_n)$ satisfying the requirement $\|X\lambda-u\|=2(h\|\lambda\|+\delta)$. Following Ref.[20], we solve this problem by the method of undetermined Lagrangian coefficients, i.e. we find vector $\lambda^\alpha$ minimizing the smoothing functional

$$M^\alpha(\lambda,u,A) = \|A\lambda - u\|^2 + \alpha\|\lambda\|^2, \tag{11}$$

and determine parameter $\alpha$ from the requirement

$$\|A\lambda^\alpha - u\| = 2\left(h\|\lambda^\alpha\| + \delta\right). \tag{12}$$

The trial values $\lambda_i$ are the solution of the system of linear algebraic equations:

$$\alpha\lambda_i + \sum_{j=1}^{n}\sum_{k=1}^{n} A_{ki}A_{kj}\lambda_j = \sum_{j=1}^{n} A_{ji}u_j, \quad i=1,\ldots,n, \tag{13}$$

obtained from the requirements of functional (11) minimum: $\left(\partial M^\alpha / \partial \lambda_j^\alpha\right)=0, j=1,\ldots,n$. At given initial approximation of regularization parameter $a$ the substitution of obtained from Eq. (13) trial sequence $\{\lambda_i\}$ to requirement (12) leads to the equation relative to parameter $\alpha$, which solution by numerical methods allows to determine its "true" value. Then solution of Eq. (13) at the found value of parameter $\alpha$ will be finite and simultaneously desired solution of the Eq. (4) (or the system of equations (10)) relative to $\lambda^\alpha_i$.

The regularized solutions $\{\lambda^\alpha_i\}$ of the system of the equations (10) corresponding to fission rates of nuclear fuel components are presented in Table III and the normalized values of this magnitudes

$$a_i^\alpha = \frac{\lambda_i^\alpha}{\sum_i \lambda_i^\alpha}, \tag{14}$$

which meets the average contributions of fissionable isotopes to total average number of fission $a_i$ in Eq. (5) during measuring ($\Delta t=10^5$ s), are given in Table IV.

The comparison of contributions from fissionable isotopes to total number of fissions (14) and (5), which characterizing the same experiment but obtained by different methods, shows the good agreement (see Table IV).

### IV. ABOUT THE ERROR OF REGULARIZED SOLUTION

First of all note that a robust method of ill-posed problem solving (if it exists) does not guarantee the obtaining of error estimate or rate of convergence of approximate solution. It only guarantees that the approximate solution tends asymptotically to exact solution [20], but does not give the answer as close or far from exact solution this solution is. But the situation changes cardinally, when the additional information about a unknown exact solution is known. In particular, if the additional information about structure of a set that the exact solution belongs to is known, in such cases one can find an error of approximate solution.

Now let us discuss the accuracy of stable approximation to normal solution (normal pseudo-solution) of our system of equations (4) or (10). The use of regularizing method for solving of ill-posed system of linear equations has a characteristic property, which is determined by following important theorem [22]: in case of the search of normal solution $\lambda$ of combined system of equations (such as (10)) by regularizing method with the use of smoothing functional (11) such value of regularization parameter $\alpha=\alpha(h)$ exists that following expression for error estimate of approximate solution is satisfied at $0 \leq h \leq h_0 = const$ and $\sigma \geq 0$:

$$\left\|\lambda^{\alpha(h)} - \lambda\right\| \leq \frac{\left\|\overline{A}^+\right\| \cdot \left[\sigma + h(1+\sqrt{2})\|\lambda\|\right]}{1 - h\left\|\overline{A}^+\right\|} = O(h+\sigma), \tag{15}$$

where $\|\cdot\|$ is Euclidean norm.

This estimation is exact under the order of values of the errors $\sigma$, $h$ and pseudoinverse matrix $\|\overline{A}^+\|$. Thus the characteristic property consists in the fact that the approximate solution obtained by this method according approximate data of the problem such as (10) has an optimal order of accuracy (15) only in case if the system of equations is solvable (more detail see in [23]). If it is unsolvable, but its normal pseudosolution is searched, the regularizing method not always gives optimal order of accuracy. It is necessary to note that to judge about the solvability of exact system of equations by its approximate data is not possible[23].

This is not our case. The system of equations is solvable because it is *a priori* known about the existence of exact solution, which in the generalized form is equivalent to reactor thermal power rating $<W_{NPP}>$ (see Table IV) estimated by formula (5):

$$W_{NPP} = \sum_i \lambda_i{}^i E_f = \lambda \sum_i \alpha_i{}^i E_f.$$

In practice $<W_{NPP}>$ is usually determined by the independent method, for example, by calorimetric method, whose error is approximately 2 % [15]. Obtained approximate solutions $\{\lambda^{\alpha}_i\}$ of the system of equations (10) coincide with WWER-440 reactor power rating of Rovno NPP [4, 15, 16] with the error no more than 0.07 % (see Table IV). The made estimation has shown that taking into account 2% error of calorimetric method the total error of the approximate solution exceeds 2-3%. Let us show this by Eq. (15), which because of the system (10) solvability is applicable to error estimation of approximate solutions for the experiments [4, 15, 16].

For the values of energy spectrum "on detection place" $\eta(E_\nu)$, i.e. $E_\nu = 2.5$; 3.0; 3.5; 4.5 MeV, which were used for solving of the system of equations (4), the relative error of experimental data is $\sigma \sim 2.71$ % in Ref. [15]; $\sigma \sim 3.18$ % in Ref.[4]; $\sigma \sim 2.59$ % in Ref.[16] (see Table II). On the other hand, operator error $\|A_h - A\| \leq h$ in Eq. (4) is determined mainly by the average error of electron antineutrino energy spectrum $\rho(E_\nu) = \sum(\alpha_i \cdot \rho_i)$, emitted by the fission products of nuclear fuel, and comes to $h \sim 5.7$ %. Now it is necessary to take into account that the error of experimental data $\sigma$ in the absolute values greatly exceeds the average error of $h$ in Eq. (4), i.e. $\sigma \gg h$ (see Table II). It follows that in our case the optimum order of accuracy (15) is mainly determined by the order of error value $\sigma$ of experimental data of inverse problem (10):

$$\left\|\lambda^{\alpha(h)} - \lambda\right\| = \|\Delta\lambda\| \leq O(\sigma). \tag{16}$$

This, in its turn, makes it possible to estimate the of error accuracy of each component of approximate regularized solution taking into account obtained average error estimates $\sigma$ for experimental data [4, 15-16] and supposition that the errors are uniformly distributed on the components of regularized solution $\lambda = (\lambda_1, \ldots, \lambda_4)$:

$$\Delta\lambda = \Delta\lambda_i = \frac{1}{2}\|\Delta\lambda\| = \begin{cases} O(1.355\%)^{\text{Ref.15}} \\ O(1.590\%)^{\text{Ref.4}} \\ O(1.295\%)^{\text{Ref.16}} \end{cases}. \tag{17}$$

At the same time fission rates $\lambda_0$, which are a sum of the components of regularized solution $\{\lambda^\alpha_i\}$ of the systems of equations (10) for every experiment [4, 15, 16] (see Table III), make it possible to estimate the total approximate solution and its accuracy:

$$\lambda_0 = \sum_i \lambda_i^\alpha \pm \|\Delta\lambda\| = \begin{cases} 4.427 \cdot 10^{19} \pm O(2.71\%)^{\text{Ref.15}} \\ 4.181 \cdot 10^{19} \pm O(3.18\%)^{\text{Ref.4}} \\ 4.182 \cdot 10^{19} \pm O(2.59\%)^{\text{Ref.16}} \end{cases}. \quad (18)$$

Here the question arises: how to switch from estimation of optimal order of accuracy of approximate solution (18) to direct estimation of accuracy of this solution? According to the above-mentioned, a general plan of such a procedure is clear: basing on the certain model notions concerning the solution it is necessary to use the additional information, which will help to "narrow" the solution set, which exact solution belongs to, and in that way to find the grounded error estimate of selected solution set. Let us show how we did it in this case.

It is known that because of $^{235}$U burnup and plutonium accumulation the reactor antineutrino spectrum and the cross-section of the reaction (1) change in time. As a result of burnup the contributions from isotopes $\alpha_i$ change noticeable during campaign also [4]. As it was experimentally shown in Ref.[24], the spectrum changes can amount to 10%, and total cross-section of reaction (1) changes by almost 6% during one campaign. These changes are comparable with the values of the errors in Eq. (18) or even exceed them. In practice the contributions from isotopes $\alpha_i$ are usually calculated by the results of radiochemical analysis of spent nuclear fuel. In this case the error for reactor type WWER-400 is approximately 5% (relative) [24]. It means that, if the values of the contributions from isotopes $\alpha_i$, used in direct problem (Eq. (2)) and found in inverse problem (Eq. (4)) are in close agreement with all other equal conditions (see Table IV), the estimation of the optimal order of accuracy of the solution (Eqs. (16-18)) and error estimate of contributions from isotopes $\alpha_i$, obtained by radiochemical analysis of spent fuel [24] are in close agreement too. In other words, the estimation of the order of error accuracy of each components of approximate regularized solution in Eq. (4) can be written down in the acceptable for calculations form:

$$\|\lambda^{\alpha(h)} - \lambda\| = \|\Delta\lambda\| \le O(\sigma) \cong 2\sigma, \quad (19)$$

which makes clear rather abstract sense of Eqs. (16)-(18). For example, Eq. (18) under the condition (19) makes it possible to estimate the order of error accuracy of relative contributions from isotopes $\alpha_i$ (Table IV), i.e. each components of approximate regularized solution (18) of balance equation (4) for the experiments [4, 15, 16]

$$\alpha_i = \frac{\lambda_i}{\lambda_0} \Rightarrow \frac{\Delta\alpha_i}{\alpha_i} \cong \sqrt{\left(\frac{\Delta\lambda_i}{\lambda_i}\right)^2 + \left(\frac{\Delta\lambda_0}{\lambda_0}\right)^2} = \begin{cases} 5.6\%^{\text{Ref.15}} \\ 6.6\%^{\text{Ref.4}} \\ 5.3\%^{\text{Ref.16}} \end{cases}, \quad (20)$$

which fits the data of radiochemical analysis of spent nuclear fuel [24].

It is important that once found condition (19) is valid for all similar experiments and does depend both on improvement of the measurement statistics (for example, due to the increase of sampling) and vice versa.

Finally, to complete the picture let us note that there is a method, which makes it possible to evade the difficulties concerned with solvability or insolubility of the system of equations such as (10). It turned out, that the theory of discrepancy generalized principle developed in [23] for the solving of nonlinear ill-posed problems makes it possible to construct a method for the solving of ill-posed systems of linear equations with the approximate data. This method (so-called pseudo-inverse matrix method), whose accuracy does not depend on solvability or insolubility of the exact system, will be applied in our next paper for the analysis and the choice of stable and reliable procedure of the determination of accuracy of approximate solution of inverse problem of neutrino diagnostics for intrareactor processes with allowance for statistics of reactor antineutrino production and detection [25-27].

## V. CONCLUSIONS

The method of the reconstruction of nuclear density of each component of nuclear fuel and determination of their dynamics during reactor operation is developed. It is shown that the neutrino method has all necessary properties of an independent and absolute method for the remote on-line diagnostics of basic parameters of reactor core, starting with the determination of current heat power and dynamics of concentration of nuclear fuel components and ending with the identification of daughter fission products and the determination of neutron fluence.

The proposed method is a completely noncontacting method, the inspection equipment can operate as "black box" and can transmit the information by the communication satellite channels, measuring data of the neutrino detector are defied the falsification. Due to these merits the neutrino method can have also specific safeguards applications[28-29].

## AKNOWLEDGMENTS

The authors are grateful to L.A. Mikaelyan for useful discussions and attention to our research.

Table I

The experiment geometry and basic characteristics of detector and reactor

|  | Experiment | | |
|---|---|---|---|
|  | Afonin et al.[15] | Klimov et al.[4] | Kopeykin et al.[16] |
| Target | liquid scintillator ($C_nH_{2n}$) | | |
| $N_p$ ($10^{28}$) | 1.506±1.5 % | 5.820±1.5 % | 5.820±1.5 % |
| $\varepsilon_0$ | 0.322±5.5 % | 0.321±5.5 % | 0.321±5.5 % |
| $\gamma$ | 1.00 | 0.75 | 0.75 |
| $<R^2>^{1/2}$, m | 18.18±0.3 % | 18.00±0.3 % | 18.00±0.3 % |
| $<W>$, MW | 1452±2.0 % | 1375±2.0 % | 1375±2.0 % |
| $\Delta T_{e+}$, MeV | 0.50 | 0.55 | 0.60 |
| $n_\nu$ during $10^5$ c | 309.5±0.7 % | - | 835.0±0.3 % |
| Number of measurements | 66 | 70 | 174 |

Table II

Experimental η($E_ν$) and converted $ρ_i$ energy spectra in the 2-9 MeV range

| $E_ν$, MeV | η($E_ν$), (MeV)$^{-1}$ | | | $ρ_i$, (MeV×fission)$^{-1}$ | | | |
|---|---|---|---|---|---|---|---|
| | η(E) [15] | η(E)/0.3 [4] | η(E)/0.3 [16] | $^{235}$U | $^{239}$Pu | $^{238}$U | $^{241}$Pu |
| 2.0 | 23.032±3.6 | 42.857±2.85 | 21.285±2.21 | 0.130(+1)±4.2 | 0.107(+1)±4.5 | 0.153(+1) | 0.124(1)±4.3 |
| 2.5 | 57.928±3.6 | 73.571±2.76 | 53.285±2.23 | 0.900(0)±4.2 | 0.710(0)±4.3 | 0.111(+1) | 0.870(0)±4.0 |
| 3.0 | 79.091±2.4 | 80.357±3.83 | 74.143±2.40 | 0.673(0)±4.2 | 0.491(0)±4.3 | 0.835(0) | 0.623(0)±4.0 |
| 3.5 | 87.267±2.4 | 91.428±2.91 | 81.357±3.30 | 0.473(0)±4.2 | 0.317(0)±4.3 | 0.586(0) | 0.420(0)±3.9 |
| 4.0 | 84.077±3.4 | 82.857±5.95 | 73.042±3.11 | 0.283(0)±4.2 | 0.190(0)±4.4 | 0.386(0) | 0.270(0)±3.9 |
| 4.5 | 71.503±2.3 | 78.428±3.11 | 60.714±2.28 | 0.172(0)±4.2 | 0.107(0)±4.8 | 0.245(0) | 0.157(0)±4.2 |
| 5.0 | 62.129±3.2 | 60.000±4.13 | 53.357±4.73 | 0.105(0)±4.2 | 0.576(-1)±5.2 | 0.152(0) | 0.920(-1)±4.4 |
| 5.5 | 49.378±2.2 | 52.142±3.17 | 35.757±2.18 | 0.617(-1)±4.2 | 0.350(-1)±5.9 | 0.908(-1) | 0.525(-1)±4.9 |
| 6.0 | 35.377±3.2 | 37.857±3.75 | 25.500±1.91 | 0.370(-1)±4.3 | 0.177(-1)±6.8 | 0.549(-1) | 0.267(-1)±5.6 |
| 6.5 | 26.007±2.7 | 22.499±2.63 | 15.714±1.93 | 0.203(-1)±4.4 | 0.940(-2)±7.4 | 0.328(-1) | 0.139(-1)±6.1 |
| 7.0 | 16.744+2.7 | 14.999±2.67 | 7.428±1.73 | 0.105(-1)±4.7 | 0.468(-2)±11 | 0.176(-1) | 0.683(-2)±7.0 |
| 7.5 | 8.693±0.48 | 9.285±1.85 | 2.642±1.83 | 0.429(-2)±5.0 | 0.180(-2)±19 | 0.787(-2) | 0.254(-2)±8.0 |
| 8.0 | 3.562±0.42 | 6.071±2.31 | 0.134+0.56 | 0.136(-2)±6.0 | 0.500(-3)±35 | 0.313(-2) | 0.890(-3)±11 |
| 8.5 | 1.033±0.45 | 2.142+1.82 | - | 0.237(-3)±10 | 0.220(-3)±80 | 0.526(*-3) | 0.235(-3)±24 |
| 9.0 | 0.139±0.36 | 1.428+1.82 | - | 0.560(-4)±27 | 0.398(-4) | 0.124(-3) | 0.470(-4)±90 |

Notes:

a) errors are given in %;

b) for the converted spectra $ρ_i$ decimal exponent is shown in brackets, errors are given in % with 90 % confidence level.

Table III

Regularized solutions $\{\lambda^{\alpha}_i\}$ of the system of equations (10)

| $\lambda \times 10^{19}$, fission $\times$ s$^{-1}$ | Afonin et al.[15] | Klimov et al.[4] | Kopeykin et al.[16] |
|---|---|---|---|
| $\lambda_5$ | 2.688 | 2.486 | 2.457 |
| $\lambda_9$ | 1.223 | 1.203 | 1.232 |
| $\lambda_8$ | 0.320 | 0.296 | 0.298 |
| $\lambda_1$ | 0.196 | 0.196 | 0.195 |
| $\lambda_0 = \sum \lambda_i$ | 4.427±2.59 % | 4.181±3.18 % | 4.182±2.71 % |

Table IV

The contributions from fissionable isotopes $a_i$ to total fission number $\lambda$

|  | Our results | Afonin et al.[15] | Our results | Klimov et al.[4] | Our results | Kopeykin et al.[16] |
|---|---|---|---|---|---|---|
| $a_5$ | 0.607 | 0.606 | 0.597 | 0.593 | 0.588 | 0.586 |
| $a_9$ | 0.276 | 0.274 | 0.288 | 0.286 | 0.295 | 0.292 |
| $a_8$ | 0.072 | 0.074 | 0.070 | 0.075 | 0.071 | 0.075 |
| $a_1$ | 0.044 | 0.046 | 0.047 | 0.047 | 0.047 | 0.047 |
| $W_{NPP}$, MW | 1452.7 | 1452 | 1374.6 | 1375 | 1375.6 | 1375 |

**FIGURE CAPTIONS**

FIG. 1. Experimental (o) and calculated ( —— ) positron spectra for the reaction (1) apart 18 m [15]. A step of the histogram is 300 keV.

FIG. 2. Experimentally measured spectra [4]: sum of the effect and correlated background (a) and correlated background (b).

FIG. 3. The measured (•) and approximated (---) positron spectra for the reaction (1) (a) and events of correlated background (b) [16]. A step of the histogram is 290 keV.

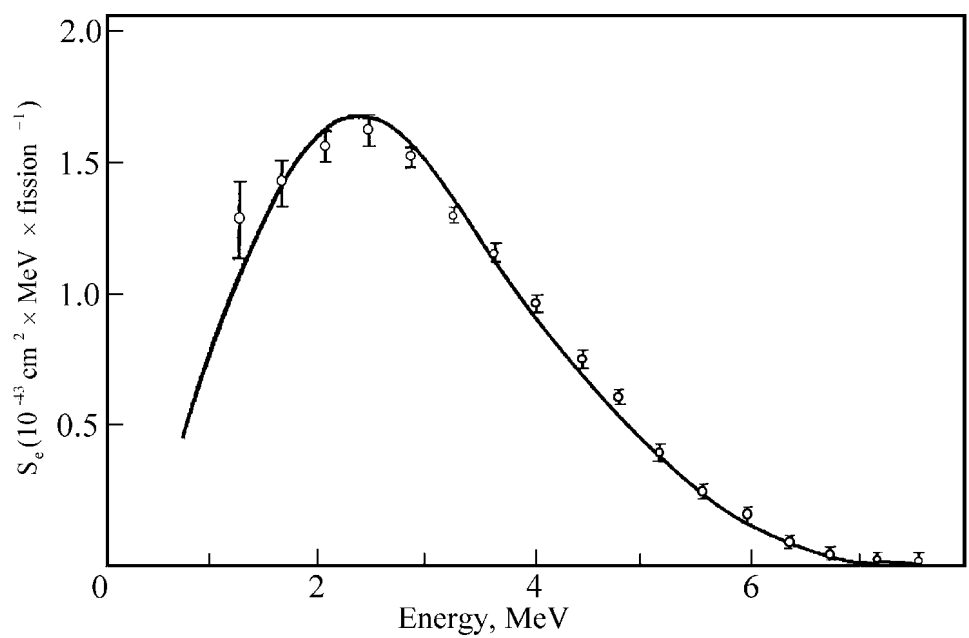

Rusov V. et al. Fig. 1

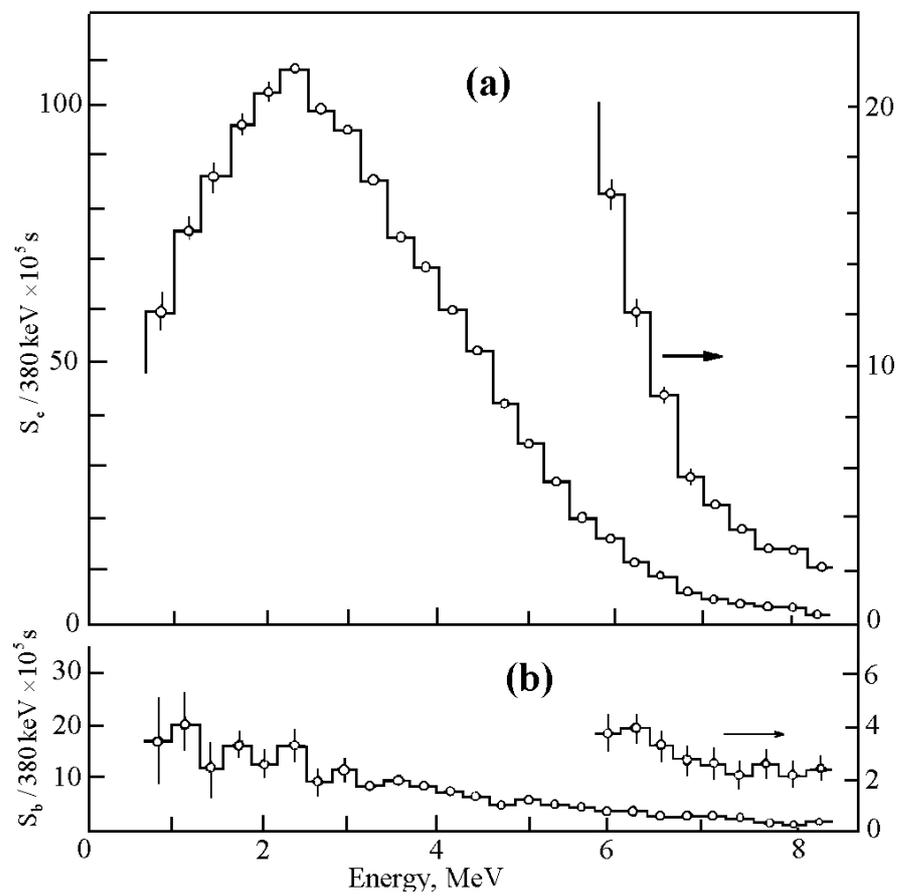

Rusov V. et al. Fig. 2

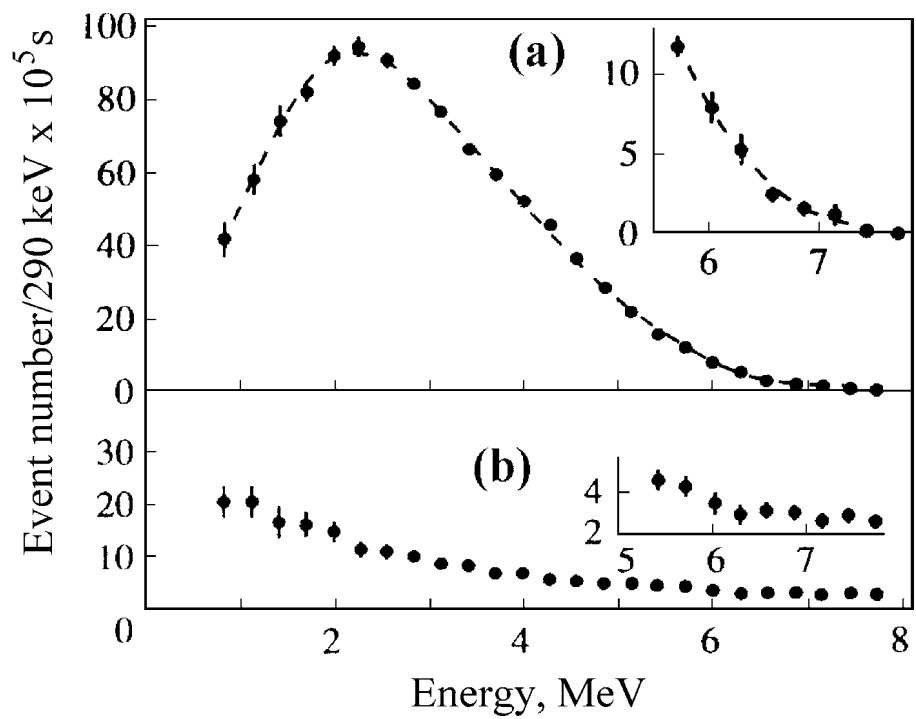

Rusov V. et al. Fig. 3